# Controllable synthesis of calcium carbonate with different geometry: comprehensive analysis of particles formation, their cellular uptake and biocompatibility

Hani Bahrom,[1,2,=] Alexander A. Goncharenko,[3,=] Landysh I. Fatkhutdinova,[4] Oleksii O. Peltek,[4] Albert R. Muslimov,[5,6] Olga Yu. Koval,[7] Igor E. Eliseev,[5] Andrey Manchev,[1,2] Dmitry Gorin,[8] Ivan I. Shishkin,[4] Roman E. Noskov,[1,2] Alexander S. Timin,[3,9,*] Pavel Ginzburg,[1,2] Mikhail V. Zyuzin[4,*]

[1]Department of Electrical Engineering, Tel Aviv University, Ramat Aviv, Tel Aviv 69978, Israel

[2]Light-Matter Interaction Centre, Tel Aviv University, Tel Aviv, 69978, Israel

[3]Peter The Great St. Petersburg Polytechnic University, Polytechnicheskaya str. 29, St. Petersburg, 195251, Russia

[4]Faculty of Physics and Engineering, ITMO University, Lomonosova 9, 191002 St. Petersburg, Russia

[5]Nanobiotechnology Laboratory, St. Petersburg Academic University, 194021 Saint Petersburg, Russia

[6]I. P. Pavlov State Medical University of St. Petersburg, Lev Tolstoy Street, 6/8, 197022 St. Petersburg, Russia

[7]Laboratory of Renewable Energy sources, St. Petersburg Academic University, 194021 Saint Petersburg, Russia

[8]Skolkovo Institute of Science and Technology, 3 Nobelya str., 121205 Moscow, Russia;

[9]Research School of Chemical and Biomedical Engineering, National Research Tomsk Polytechnic University, Lenin Avenue 30, 634050 Tomsk, Russia

*corresponding authors: a_timin@mail.ru, timin@tpu.ru, mikhail.zyuzin@metalab.ifmo.ru

= equal contribution

**Abstract**

Carefully designed micro- and nanocarriers can provide significant advantages over conventional macroscopic counterparts in biomedical applications. The set of requirements including a high loading capacity, triggered release mechanisms, biocompatibility, and biodegradability should be considered for the successful delivery realization. Porous calcium carbonate ($CaCO_3$) is one of the most promising platforms, which can encompass all the beforehand mentioned requirements. Here, we study both the particles formation and biological applicability of $CaCO_3$. In particular, anisotropic differently shaped $CaCO_3$ particles were synthesized using green sustainable approach based on co-precipitation of calcium chloride and sodium carbonate/bicarbonate at different ratios in the presence of organic additives. The impact of salts concentrations, reaction time, as well as organic additives was systematically researched to achieve controllable and reliable design of $CaCO_3$ particles. It has been demonstrated that the crystallinity (vaterite or calcite phase) of particles depends on the initial salts' concentrations. The loading capacity of prepared $CaCO_3$ particles is determined by their surface properties such as specific surface area, pore size and zeta-potential. Differently shaped $CaCO_3$ particles (spheroids, ellipsoids, toroids) were used to evaluate their uptake efficiency on the example of C6 glioma cells. The results show that the ellipsoidal particles possess a higher probability for internalization by cancer cells. All tested particles were also found to have a good biocompatibility. The capability to design physicochemical properties of $CaCO_3$ particles has a significant impact on drug delivery applications, since the particles geometry substantially affects cell behavior (internalization, toxicity) and allows outperforming standard spherical counterparts.

**Keywords:** Calcium carbonate, formation mechanisms, drug carriers, adsorption, cell uptake, toxicity.

**Synopsis:** Calcium carbonate particles with controllable geometries have been synthesized using green sustainable approach and tested for drug delivery applications.

# 1. Introduction

Nano- and microparticles are extensively used as a delivery platform for biologically active compounds. Controllable physicochemical properties, enhanced loading capacity, high stability in biological fluids, feasibility of incorporation of hydrophilic and hydrophobic compounds and sustained drug release are among the functions, which particles can offer.[1–5] Compared to conventional polymer or lipid carriers, the unique characteristics of inorganic nanoparticles make them to hold a great potential for theranostic applications in both disease therapy and simultaneous bioimaging.[5]

Among widely used inorganic materials, calcium carbonate ($CaCO_3$) has broad biomedical utilizations owing to its availability, low cost, safety, biocompatibility, pH-sensitivity and biodegradability.[6] The $CaCO_3$ has three natural forms – vaterite, calcite and aragonite, they are crystalline forms of $CaCO_3$, while the first one is a metastable type of polycrystal.[7] Vaterite has drawn a great attention of scientific community owing to its peculiar optical and biochemical properties.[8,9] The beneficial properties of vaterite include its porosity and metastability along with practically relevant straightforward and low cost self-assembly synthesis. Therefore, vaterite particles can be efficiently used as biocompatible containers for drug delivery of therapeutic relevant compounds into living cells and tissues.[10] Moreover, $CaCO_3$ particles can be successfully employed as templates for the synthesis of the polymer hollow capsules made with layer-by-layer method, which are commonly used as drug delivery carriers.[11–14]

It was reported that the crystallization process of $CaCO_3$ occurs through the formation and further transformation of amorphous $CaCO_3$ into vaterite and then calcite.[15] The spherulitic growth of vaterite crystals was demonstrated to be replaced by an Oswald ripening process of crystal growth. The subsequent ripening of vaterite and formation of calcite is determined by a dissolution-precipitation process.[10] Thermodynamic stability of vaterite and aragonite is significantly lower than of calcite, which is stable ground state of vaterite and aragonite.[16] Under normal conditions vaterite particles undergo phase transition to calcite form in aqueous solutions.[17] Such transition naturally occurs over time, since recrystallization in solution provides possible kinetic paths for vaterite crystals transformation into more energetically stable forms of the calcium carbonate.[17,18] One can control the crystallinity of the resulted $CaCO_3$ particles, as well as their size, shape and porosity, by changing the synthetic conditions, such as pH values, reagents ratio,[19,20] temperature,[21,22] viscosity of the reaction,[23,24] or by adding organic and inorganic additives.[25–27] Also, the synthesis of $CaCO_3$ particles of different phases (vaterite, calcite and aragonite) with control over size and morphology can be performed without additives using a vortex fluidic device.[28] Boulos et al. investigated the influence of high shear forces on the phase behavior of $CaCO_3$.

There are a number of works describing crystallization of $CaCO_3$ in different solvents, e.g., low molecular weight alcohols.[29,30] Viscosity of the reaction significantly influences the crystallization of $CaCO_3$ due to the reduction of diffusion velocity of calcium and carbonate ions. It allows controlling the velocity of nucleation and $CaCO_3$ polycrystalline growth what leads to the size reduction of the synthesized particles. Water-soluble polyethylene glycol, ethylene glycol or glycerin are considered as the most suitable organic solvents.[13]

The temperature and water content of the co-precipitation reaction affect the phase composition of the resulted $CaCO_3$ particles. Lowering the reaction temperature leads to the formation of calcite particles.[18] At room temperature the reaction product is usually formed in the vaterite polymorph,[31] whereas further increase of the temperature induces the formation of aragonite particles.[32] The effect of the solvent composition and reaction time is related to the re-crystallization process. Low water content or water absence results in a stabilization of crystalline phase of $CaCO_3$ particles, which are formed at the early stages of the synthesis. Therefore, the introduction of organic additives such as ethylene glycol at amount higher than 80% in volume leads to the formation of vaterite and calcite phases of $CaCO_3$ particle in different ratios.[20,30]

The acidity of the reaction plays a crucial role in the formation of $CaCO_3$ particles as well, and is determined by the concentration of carbonate ($CO_3^{2-}$) ions. Carbonate ion is an anion of dibasic weak carbonic acid. This means that the constants of the acid dissociation of this compound are several orders of magnitude lower than one, namely pKa1 = 6.35 and pKa2 = 10.33. This is due to the fact that carbonate ions exist in various forms at different pH values.[33]

The rate of $CaCO_3$ formation is accelerated in the presence of increased concentration of $CO_3^{2-}$ ions, in particular, at pH>10.5. The crystallization reaction can be controlled (i) at the defined pH values of initial salts (precursors), or (ii) by changing the $CO_3^{2-}$ ions source, e.g., employing carbonate, bicarbonate ($HCO_3^-$) ions, bubbling of carbon dioxide or its absorption from the gas-saturated atmosphere.[34] The use of $CO_2$ solution allows to achieve a minimal reaction rate, however, this method is challenging due to its poor reproducibility.

Finally, the ratios of the initial salts used to produce $CaCO_3$ particles affect geometrical properties (size, shape) and surface functionalities (charge and hydrophilicity) of the resulted particles. Recently, it was shown that a high concentration of $CO_3^{2-}$ ions results in the formation of anisotropic rhomboidal and ellipsoidal geometries, while a low concentration leads to the formation of isotropic spherical particles.[20,35] In spite of a number of studies on synthesis of $CaCO_3$ particles of different geometries (e.g. toroids,[36,37] hollow spherical particles,[38,39] microsponges,[40] snowflakes,[41] flowers,[42,43]), the relation between the particles morphology (size, shape, surface properties and so forth) and their phase transformation mechanisms is rarely presented. In particular, the knowledge of mechanism of $CaCO_3$ particles formation enables their further application in various fields, e.g. drug delivery.[15,44] In regard of biomedical applications, the biocompatibility of the drug carrier platforms should be taken into account to avoid undesired side effects.[1,45]

In this work, the comprehensive and systematic studies on the formation of differently shaped $CaCO_3$ particles were performed. In particular, the influence of various concentrations of initial reagents, their ratio, reaction time, as well as organic additives on $CaCO_3$ particles formation was investigated. This study also provides structural insights of the formed $CaCO_3$ particles verified with scanning electron microscopy (SEM), laser scanning confocal microscopy (CLSM), X-ray diffraction (XRD), Brunauer–Emmett–Teller analysis (BET), thermogravimetric analysis (TGA) and differential scanning calorimetry (DSC). The formation and growth of $CaCO_3$ particles of different morphologies are discussed in details. Also, the impact of particles geometry on the cellular uptake in qualitative and quantitative manners was examined. Additionally, the biocompatibility of designed particles was tested on an example of C6 glioma cells. The schematic illustration of the performed synthetic steps is depicted in the **Figure 1**.

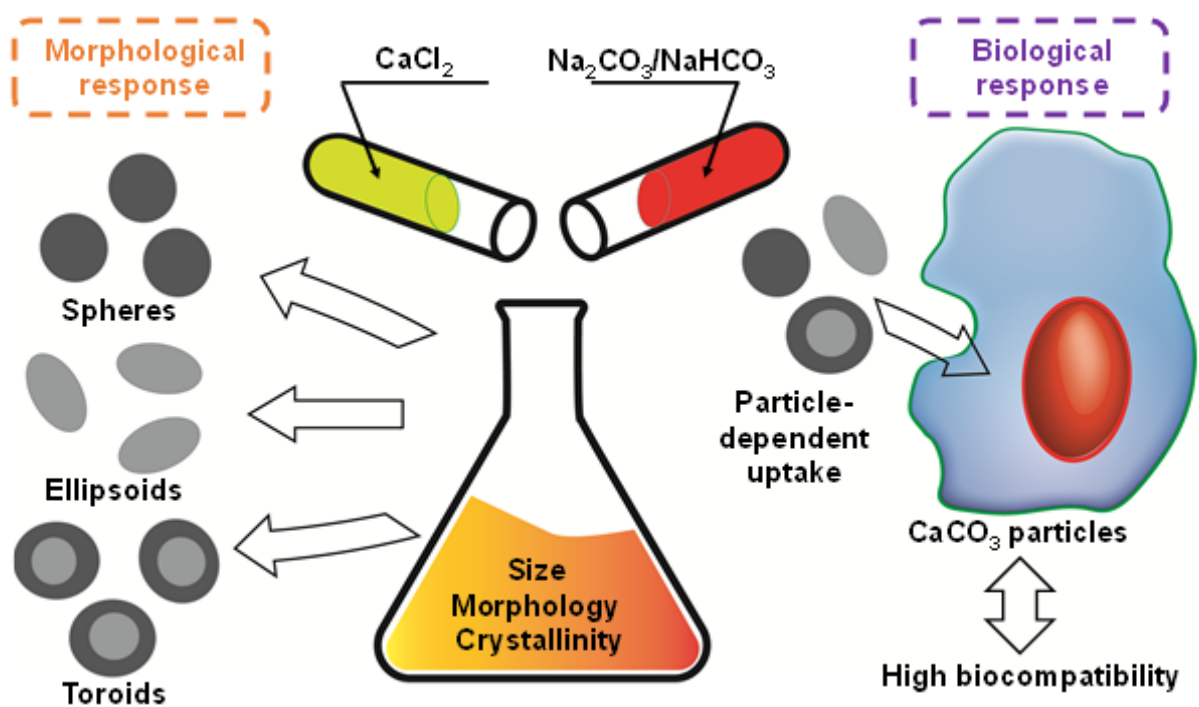

*Figure 1. Schematic illustration of the main performed steps for studying the process of differently shape CaCO₃ particles formation, as well as toxicity and uptake studies.*

## 2. Experimental section

### *2.1 Study of CaCO₃ particles formation*

Synthesis of CaCO$_3$ particles was performed in co-precipitation reaction. To study the particles formation, the different concentration of salts (CaCl$_2$, Na$_2$CO$_3$ or NaHCO$_3$), salts ratios, reaction times and introduction of additives were varied. The details of particle synthesis and the full dataset are given in the ***Supporting Information***.

### *2.2 Structural characterization of particles*

Structural characterization and fraction calculations were performed with X-ray diffractometry via measurement of 2D XRD patterns. The data were collected on a Kappa Apex II diffractometer (Bruker AXS) using Cu Kα (λ = 1,5418Å) radiation generated by a IμS microfocus X-ray tube. 2D images were converted to Θ-2Θ scans using Dioptas software [http://dx.doi.org/10.1080/08957959.2015.1059835]. The crystallite size and phase fraction were calculated by Rietveld refinement ____ref using FullProf software package.

The particles' morphology, structure, size distribution, were determined with scanning electron microscopy (SEM, Quanta200, FEG) at 10kV acceleration voltage, confocal laser scanning microscopy (CLSM, Carl Zeiss LSM 710). The details of particle synthesis and the full dataset are given in the ***Supporting Information***.

### *2.3 Surface area characterization of particles*

The Brunauer–Emmett–Teller (BET) specific surface area was calculated using adsorption data. The pore size distribution curves were calculated from the analysis of the adsorption branch of the isotherm using the Barrett–Joyner–Halenda algorithm. Errors in determining of BET surface areas and pore volumes were estimated to be within 5%.

### *2.4 Thermogravimetric analysis (TGA) and differential scanning calorimetry (DSC)*

Thermogravimetric analysis was performed on a TG 209F1 ("Netzsch", Germany). DSC was carried out using a DSC 204F1 system ("Netzsch", Germany). The experiments were performed using platinum crucibles in argon atmosphere with a heating rate of 10 ˚C min$^{-1}$. The accuracy of sample mass measurement was 1,106 g and accuracy in temperature measurement was 0.1 ˚C.

*2.5. Dynamic Light Scattering (DLS) and zeta-potential measurements*

The zeta potential of particles was measured by DLS which was performed using a Malvern Zetasizer Nano Series running DTS software and operating a 4 mW He-Ne laser at 633 nm. Analysis was performed at an angle of 173° and a constant temperature of 25 °C.

*2.6 Adsorption capacity of differently shaped $CaCO_3$ particles*

The adsorption of dye (TRITC) onto the $CaCO_3$ particles surface was measured with multiplate reader (CLARIOstar®, BMG LABTECH). For this purpose, differently shaped particles were incubated with aqueous solution of TRITC at different concentrations for 2 hours. The adsorption capacity was then calculated and plotted versus used TRITC concentration. The details of particle synthesis and the full dataset are given in the ***Supporting Information***.

*2.7 Toxicity studies*

Cell viability was measured with LIVE/DEAD assay. For this, cells were fluorescently stained with Calcein AM (for live cells) and Propidium Iodide (for dead cells). Fluorescence images were then taken with the confocal microscopy and analyzed with FIJI open source image analysis software. Detailed protocols are presented in the ***Supporting Information***.

*2.8 Uptake studies*

Internalization of differently shaped $CaCO_3$ particles was evaluated with CLSM. To this end, TRITC labelled particles were incubated with C6 glioma cells. Next day, the cytoskeleton of the cells was stained with phalloidin conjugated with AlexaFluor 488 and the cell nuclei with Propidium Iodide. Afterwards, stained C6 glioma cells were scanned with the confocal microscope using Z-stack option. An indication for intracellular localization of particles was red signal coming from fluorescently labelled particles surrounded with the green signal coming from cell cytoskeleton. The amount of internalized cells was then counted and the frequency $f(x)$ histogram of cells with have internalized x particles per cell was plotted. Additionally, the cumulative probability $p(x)$ was also plotted. Detailed protocols are presented in the ***Supporting Information***.

3. **Results and discussion**

*3.1 Calcium carbonate particles formation*

The initial effort of this work was devoted to understand the influence of the salt concentrations on $CaCO_3$ particles formation. For this purpose, particles were synthesized by mixing calcium chloride ($CaCl_2$) with sodium carbonate ($Na_2CO_3$) at the equal molar ratio (1:1) and varying the initial salt concentrations ($CaCl_2$ and $Na_2CO_3$) from $5 \times 10^{-4}$ to $5 \times 10^{-2}$ M. The reaction time was 5 min, what means the time between the start of the reaction and washing steps. In total, 3 reactions were performed: **reaction α1:** $c(CaCl_2)= 5 \times 10^{-4}$ M, $c(Na_2CO_3)= 5 \times 10^{-4}$ M; **reaction α2:** $c(CaCl_2)= 5 \times 10^{-3}$ M, $c(Na_2CO_3)= 5 \times 10^{-3}$ M; **reaction α3:** $c(CaCl_2)= 5 \times 10^{-2}$ M, $c(Na_2CO_3)= 5 \times 10^{-2}$ M. The reaction conditions are enlisted in the **Table S1**. The size and the morphology of the obtained products were further evaluated with scanning electron microscopy (SEM) (**Figure 2**). As shown in **Figure 2**, the increase of the salt concentration leads to the particle size growth: the product of the **reaction α1** consisted of amorphous $CaCO_3$ aggregates. The product of the **reaction α2** was slightly elongated quasi-spheroidal particles with the size of 519±89 nm. Finally, the product of the **reaction α3** was polydispersed particles with the sizes ranged from 500 nm to 1800 nm. The histograms of particle size distributions are shown in the *Supporting Information* (**Figure S1A**). By analyzing the **reactions α1 - α3,** we can conclude that the **reaction α2** is more preferable in terms of synthesis of spherical $CaCO_3$ particles with narrow size distribution and similar morphology. Therefore, the range of $10^{-3}$ M was chosen as an optimal value and used in further experiments.

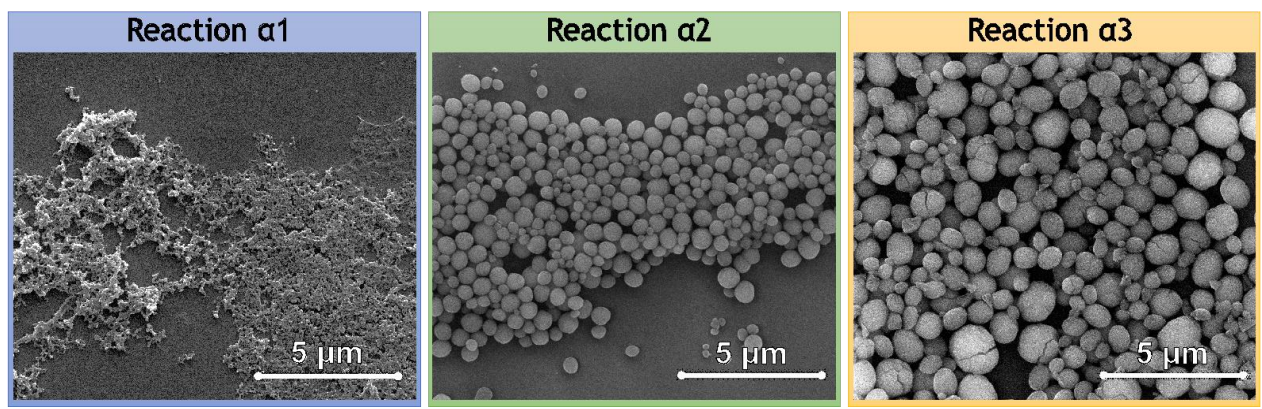

*Figure 2. Influence of salts concentrations on $CaCO_3$ particles formation: SEM images of the products of the reaction α1: $c(CaCl_2)= 5 \times 10^{-4}$ M, $c(Na_2CO_3)= 5 \times 10^{-4}$ M; reaction α2: $c(CaCl_2)= 5 \times 10^{-3}$ M, $c(Na_2CO_3)= 5 \times 10^{-3}$ M; reaction α3: $c(CaCl_2)= 5 \times 10^{-2}$ M, $c(Na_2CO_3)= 5 \times 10^{-2}$ M.*

Next, we investigated the process of $CaCO_3$ particles formation, depending on the reaction time and ion excess concentration. For this purpose, the synthesis was performed at various molar ratios of salts at the reaction time ranging from 5 to 1440 min. To understand the influence of ion excess, the following $CaCl_2:Na_2CO_3$ ratios were used: 5:1, 1:1 and 1:5, and resulted in **reaction β1:** $c(CaCl_2)=25 \times 10^{-3}$ M, $c(Na_2CO_3)=5 \times 10^{-3}$ M (5:1); **reaction β2:** $c(CaCl_2) = 5 \times 10^{-3}$ M, $c(Na_2CO_3) = 5 \times 10^{-3}$ M (1:1); **reaction β3:** $c(CaCl_2) = 5 \times 10^{-3}$ M, $c(Na_2CO_3)= 25 \times 10^{-3}$ M (1:5). Note, that **reaction β2** was not a direct replica of **reaction α2**, since the latter did not include particles growth formation.

The SEM images of the resulted products (**reactions β1-β3**) demonstrate the process of CaCO$_3$ particles formation over time (**Figures 3, S2**). At the molar ratios of CaCl$_2$:Na$_2$CO$_3$ = 1:1 and 5:1, the CaCO$_3$ particles with the sizes of 500 - 700 nm were observed after 30 min of the reaction. The continuous increase in particle size for both ratios was detected. In the case of molar ratio of 1:5, the particles formation occurred already after 10 min and the size of particles continued to grow up to ≈ 1 μm. As shown in **Figure 3B**, the excess of the CO$_3^{2-}$ ions accelerates the reaction rate and results in the formation of the CaCO$_3$ particles on early stages. For the additional verification XRD analysis of the products from the **reaction β1** and **reaction β3** with the reaction time 20 min was performed (**Figure S8**). According to the XRD scans, after 20 min of the reaction **β1** the large fraction of amorphous material was detected, while in case of **reaction β3** the peak that indicated the amorphous phase was absent. These results are in agreement with the SEM analysis and reveal the faster formation of crystalline CaCO3 particles in the **reaction β3**. For all tested molar ratios (5:1, 1:1 and 1:5), the stable CaCO$_3$ particles were formed after 1440 min of co-precipitation reaction. Besides the process of particles formation as a function of time, the excess of Ca$^{2+}$ or CO$_3^{2-}$ ions have influence on particles geometry. To evaluate it, the ellipticity parameter was used. The analysis of the particles ellipticity was performed after 1440 min of co-precipitation reaction. As shown in **Figure 3C**, at molar ratio of CaCl$_2$:Na$_2$CO$_3$ = 5:1 (**reaction β1**), the spherical particles were formed. However, the further increase in the concentration of CO$_3^{2-}$ ions resulted in the formation of prolate ellipsoids (particles ellipticity < 1) (**Figure 3C**). As seen from **Figure 3B, C**, the excess of Ca$^{2+}$ ions (**reactions β1**, Ca$^{2+}$/CO$_3^{2-}$ = 5:1) slows down the process of particles formation and promotes the growth of the spheroidal particles.

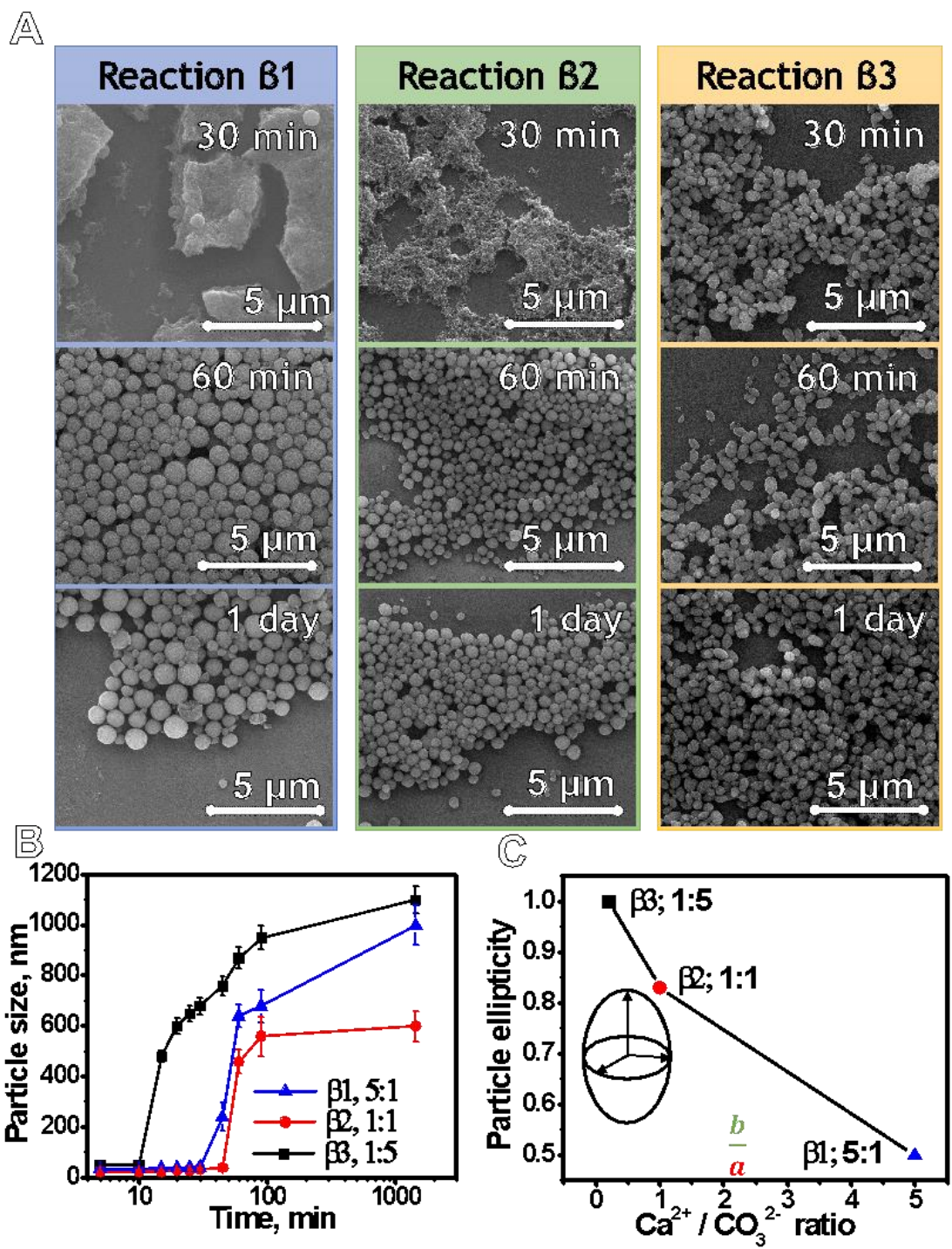

*Figure 3. Study of CaCO$_3$ particles formation kinetics using CaCl$_2$ and Na$_2$CO$_3$: **A**. SEM images of the products of the **reaction β1** (30 min, 60 min and 1 day), the **reaction β2** (30 min, 60 min and 1 day), the **reaction β3** (30 min, 60 min and 1 day). **B**. Dependence of CaCO$_3$ particles diameter on the reaction time as determined from SEM images. Data are presented as mean ± standard deviation. **C**. Dependence of particle ellipticity on the Ca$^{2+}$/CO$_3^{2-}$ ratio.*

Previous studies revealed that the introduction of HCO$_3^-$ ions reduces the reaction rate.[27] In this study, the NaHCO$_3$ was used to evaluate the influence of HCO$_3^-$ ions.[33,34] The following **reactions** γ were performed: **reaction γ1**: c(CaCl$_2$) = 25x10$^{-3}$ M, c(NaHCO$_3$) = 5x10$^{-3}$ M; **reaction γ2:** c(CaCl$_2$) = 5x10$^{-3}$ M, c(NaHCO$_3$) = 5x10$^{-3}$ M; **reaction γ3**: c(CaCl$_2$) = 5x10$^{-3}$ M, c(NaHCO$_3$) = 25x10$^{-3}$ M; **reaction γ4**: c(CaCl$_2$) = 5x10$^{-3}$ M, c(NaHCO$_3$)=50x10$^{-3}$ M; **reaction γ5**: c(CaCl$_2$)=25x10$^{-3}$ M, c(NaHCO$_3$) = 5x10$^{-3}$ M. All reaction conditions are enlisted in **Table S1**.

In the **reactions γ1-γ3** the same conditions were used as for the reactions **β1-β3**, but NaHCO$_3$ was used instead of Na$_2$CO$_3$. Additionally, the **reactions γ4-γ5** with increased concentration of HCO$_3^-$ were performed to track the particles formation and their geometry. The reaction time ranged from 5 to 1440 min. The growth of the CaCO$_3$ particles as function of time was retrieved from SEM images (**Figure 4**). As expected, the obtained data reveal that the use of HCO$_3^-$ reduces the rate of the particles growth compared to the **reactions β1-β3**. The formation of CaCO$_3$ particles was observed after 90 min of the reaction for all tested molar ratios (5:1, 1:1, 1:10, 1:15) except 1:5. The obtained particle sizes ranged from 400 to 650 nm after 1440 min. At the molar ratio of 1:5 (**reaction γ3**), the process of the particles growth was similar to the **reaction β3**. In the **reaction γ3,** spheroidal particles were already formed after 15 min of the reaction. In case of **reactions γ1, γ2,** the excess of Ca$^{2+}$ ions reduces the rate of particles formation, what is in agreement with the previous observations (**Figure 3B**). Additionally, the HCO$_3^-$ ions postpone the particle formation. Thus, two factors such as the excess of Ca$^{2+}$ and presence of HCO$_3^-$ ions reduce the rate of particles growth.

The product of the **reaction γ3** was the particles with heterogeneous morphology containing spheroids and ellipsoids (**Figure 4A**). This stems from the fact that at the beginning of the reaction (30 and 60 min), the spherical CaCO$_3$ particles were formed due to the dissociation of HCO$_3^-$ ions into CO$_3^{2-}$ ions. The presence of dissociated CO$_3^{2-}$ ions with higher concentration of Ca$^{2+}$ ions reproduces the **reaction β1** (Ca$^{2+}$/CO$_3^{2-}$ = 5:1), which results in the formation of the spheroids. However, the further dissociation of the HCO$_3^-$ ions leads to the saturation of CO$_3^{2-}$ ions in solution and shifts the ion balance, reflecting the conditions of the **reaction β3** (Ca$^{2+}$/CO$_3^{2-}$ = 1:5), what promotes the growth of ellipsoids. Thus, the dynamic change of conditions can be responsible for the formation of particles with heterogeneous morphology.

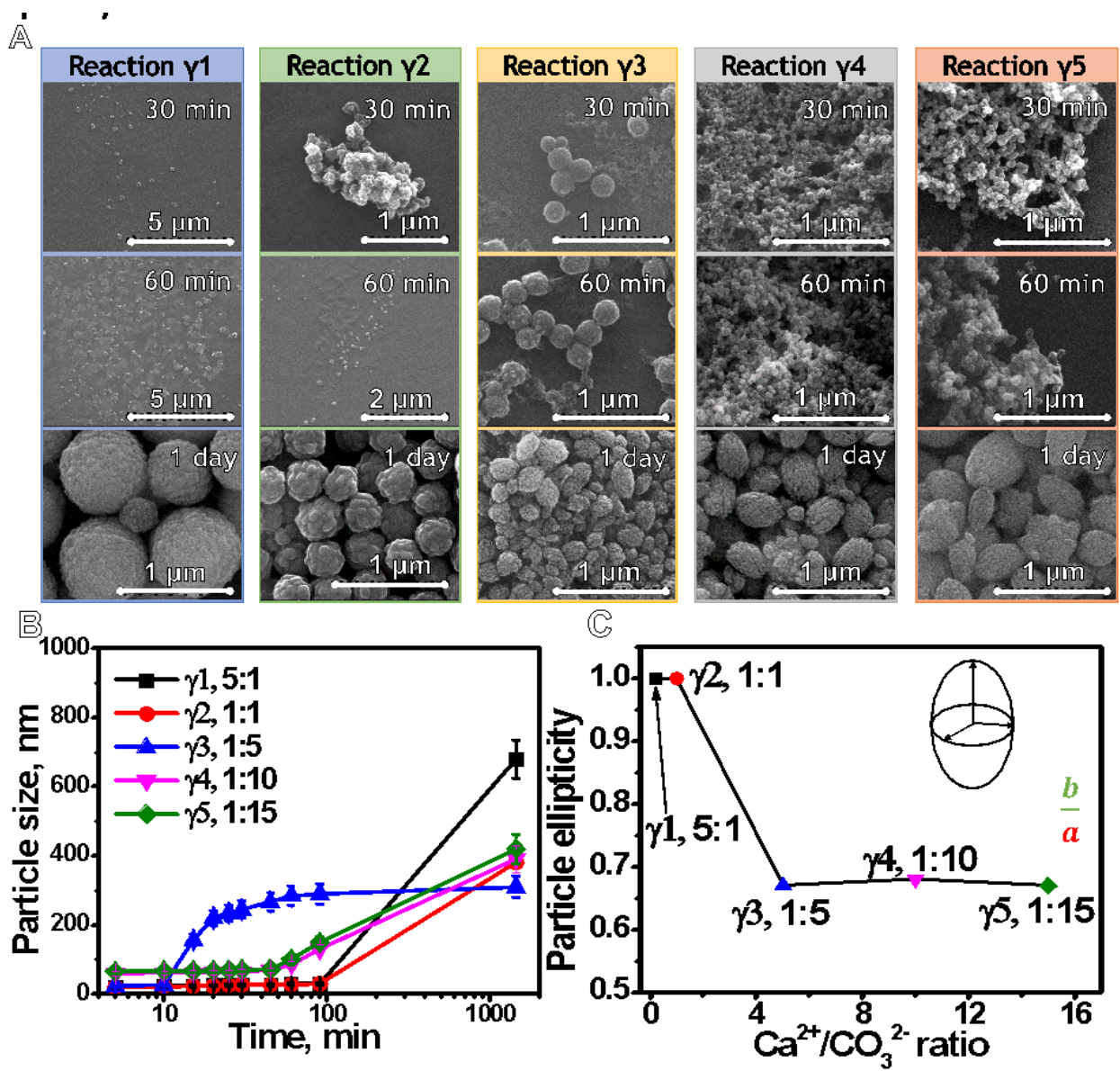

*Figure 4. Study of CaCO₃ particles formation kinetics using CaCl₂ and NaHCO₃: **A**. SEM images of the products of **reaction γ1** (30 min, 60 min and 1 day), the **reaction γ2** (30 min, 60 min and 1 day), the **reaction γ3** (30 min, 60 min and 1 day), the **reaction γ4** (30 min, 60 min and 1 day), the **reaction γ5** (30 min, 60 min and 1 day) **B**. The particle size of CaCO₃ versus the reaction time as determined from SEM images. Data are presented as mean ± standard deviation. **C**. Dependence of particle ellipticity from the $Ca^{2+}/CO_3^{2-}$ ratio.*

Among the considered **reactions γ1- γ5**, the **reaction γ3** appears to be the most convenient because of the shortest time of the particles formation. However, the obtained particles possessed heterogeneous morphology. The particles in this sample can be homogenized by varying the initial concentration of $HCO_3^-$ ions. To this end, the additional **reactions γ6** and **γ7** were performed, where the salt ratio remained the same (CaCl₂:NaHCO₃=1:5), but their initial concentrations were increased: **reaction γ6**: c(CaCl₂) = 10×10⁻³ M, c(NaHCO₃)=50×10⁻³ M; **reaction γ7**: c(CaCl₂) = 15×10⁻³ M, c(NaHCO₃) = 75×10⁻³ M. SEM images clearly demonstrate the formation of ellipsoidal particles (**Figure 5A**). According to the obtained results, in the case of the **reactions γ3** and **γ6** the particles growth has the similar reaction rate and induction time (slow stage of a chemical reaction). The **reaction γ7** occurs much faster and already after 10 min the stable $CaCO_3$ particles were formed. Thus, the increase of the concentration of $HCO_3^-$ ions resulted in the formation of homogenized ellipsoidal particles with a larger aspect ratio. To sum up, **reaction γ7** results in stable homogenized ellipsoidal particles with the shortest time of particles formation.

Additionally, the reactions at ratios of 20:1 and 1:20 were performed (reaction time was 1440 min) to prove that higher concentrations of $Ca^{+2}$ or $HCO_3^-$ have impact on the shape of formed particles. According to the obtained data (**Figure S4**), at ratio of 20:1 ($Ca^{2+}$ excess) the slow formation of spheroidal particles was observed, while at ratio of 1:20 ($HCO_3^-$ excess) the fast formation of ellipsoids was detected.

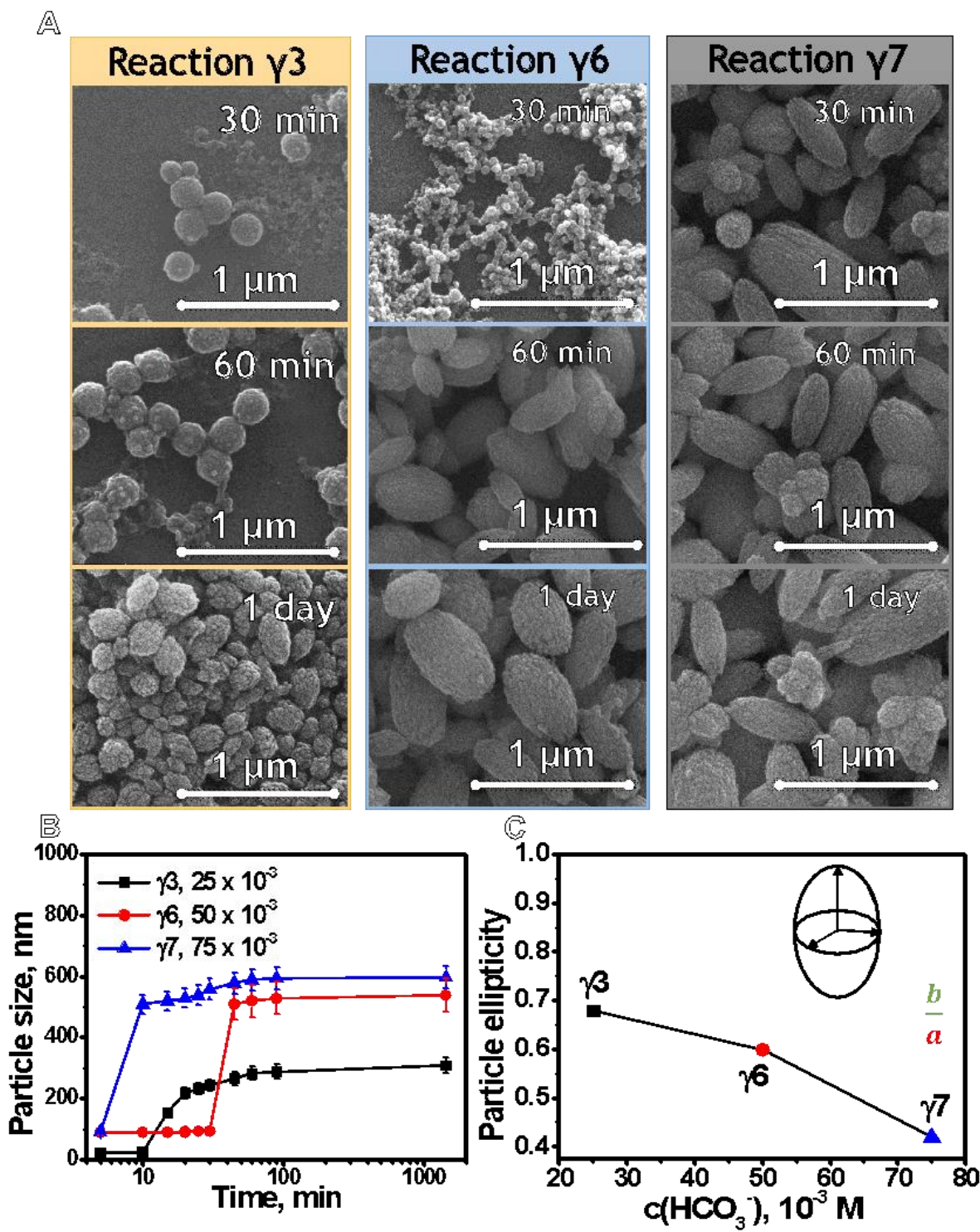

*Figure 5. Study of CaCO₃ particles formation using CaCl₂ and NaHCO₃:* ***A.*** *SEM images of the products of **reaction γ3** (30 min, 60 min and 1 day), the **reaction γ6** (30 min, 60 min and 1 day), the **reaction γ7** (30 min, 60 min and 1 day)* ***B.*** *The particle size of CaCO₃ versus the reaction time as determined from SEM images;* ***C.*** *Dependence of particle ellipticity on the $Ca^{2+}/CO_3^{2-}$ ratio. Data are presented as mean ± standard deviation.*

To evaluate the influence of molar ratios of $CaCl_2$ and $Na_2CO_3$ for **reactions β1-β3** and $CaCl_2$:$NaHCO_3$ for **reactions γ1-γ4** on the calcite /vaterite phase composition, we employed the X-ray diffraction (XRD) analysis (**Figure S7**). In case of **reaction β1,** the products composed of 55.4% of vaterite phase and 44.6% of calcite phase were formed ($CaCl_2$:$Na_2CO_3$ =5:1). The same tendency in the phase content was observed for the **reaction γ1**, where $HCO_3^-$ ions were used ($CaCl_2$:$NaHCO_3$ =5:1). In contrast, in the **reactions β2, γ2** ($CaCl_2$:$Na_2CO_3$/$NaHCO_3$ =1:1), the percentage of vaterite phase increased up to 85.6% and 84.9%, respectively. The same trend in the increase of vaterite phase was also detected for other **reactions β3, γ3, γ4**. These results indicate that the increase of $CO_3^{2-}$ or $HCO_3^-$ ions concentration is favorable for the formation of vaterite, while increased concentration of $Ca^{2+}$ ions resulted in high content of calcite phase.

## 3.2 Influence of the organic additives of the $CaCO_3$ particles formation

In order to control the size and shape of $CaCO_3$ particles formation, organic additives are often used in the co-precipitation reaction.[46] The charged polymers can interact with $Ca^{2+}$ ions forming Ca-polymer globules resulting in the formation of metastable amorphous $CaCO_3$. These metastable complexes define the geometrical parameters of final products.[47] In this study, dextran sodium sulfate (DS) and poly(styrene sulfonate) sodium (PSS) were used as macromolecular matrices. These organic additives are able to form the complexes with the $Ca^{2+}$ ions.[47–49] Based on the previous data, the 5:1 molar ratio of salts resulted in formation of spheroids (excess $Ca^{2+}$, **reaction β1**). Therefore, the same molar ratio of $CaCl_2$ and $Na_2CO_3$ (5:1) was used, but with the addition of PSS at different concentrations: **reaction δ1** [c($CaCl_2$)=25x10$^{-3}$ M, c($Na_2CO_3$)= 5x10$^{-3}$ M), c(PSS)=0.5 mg/mL], **reaction δ2** [c($CaCl_2$)= 25x10$^{-3}$ M, c($Na_2CO_3$)=5x10$^{-3}$ M], c(PSS)=1 mg/mL), **reaction δ3** [c($CaCl_2$)=25x10$^{-3}$ M, c($Na_2CO_3$)=5x10$^{-3}$ M), c(PSS)=2 mg/mL].

According to the **reactions δ1-δ3**, the particles with unique morphology were formed (**Figure 6**). The obtained particles had a toroidal-like form with a closed hole in the middle. Similar structures were previously reported, but with larger sizes (50 μm in the Ref.[36], 5 μm in the Ref.[37] and 1 μm obtained in this work). In order to simplify the obtained form of particles and make it more clear for the readers, we identified the formed $CaCO_3$ particles as toroids. The toroidal shape of obtained $CaCO_3$ particles can be explained by the directional growth of the Ca-polymer globules onto adsorbed polymers. In other words, the adsorbed polymer (PSS) is responsible for the growth of the Ca-polymer globules in the defined direction (**Figure 6, scheme**). At the highest concentration of the PSS (**reaction δ3**) polymer is adsorbed onto the $CaCO_3$ particles forming a dense film predominantly on the toroid's edges and almost absent in the middle of the formed structure. At the lowest PSS concentration (**reaction δ1**) less polymer is adsorbed onto the edges of the forming particles, and therefore, the structure of obtained toroids was less pronounced. Thus, we found out that it is possible to alter morphology of the final particles by employing various concentrations of PSS.

Next, the influence of organic additives on the formation of $CaCO_3$ particles in the excess of $CO_3^{2-}$ ions was also studied [**reaction δ4:** $(c(CaCl_2)=25 \times 10^{-3}$ M, $c(Na_2CO_3)= 5 \times 10^{-3}$ M), $c(PSS)=0.5$ mg/mL)]. Since the obtained toroids in the **reaction δ1** were more monodispersed, the concentration of PSS 0.5 mg/mL was considered as the most optimal. In the **reaction δ4**, the ellipsoidal particles were formed. This can be explained by the lack of $Ca^{2+}$ ions, which are needed to form the Ca-PSS globules. Therefore, the product of the **reaction δ4** was similar to the product of the **reaction β3** without PSS addition. Thus, the concentration of $Ca^{2+}$ ions in the co-precipitation reaction affects the geometry and morphology of final products.

In case of PSS monomers (sodium 4-vinylbenzenesulfonate, SV), the formation of toroidal-like structures was not observed, highlighting that the monomeric structure was not able to guide the particles growth in the defined direction (**reaction δ5**: $c(CaCl_2)=25 \times 10^{-3}$ M, $c(Na_2CO_3)= 5 \times 10^{-3}$ M, SV=0.5 mg/mL).

Another sulfate groups-containing polymer (dextran sodium sulfate, DS) was tested as an example of biodegradable polymer to produce toroidal $CaCO_3$ particles (**reaction δ6**: $c(CaCl_2)=25 \times 10^{-3}$ M, $c(Na_2CO_3)= 5 \times 10^{-3}$ M, DS=0.5 mg/mL). According to the obtained results, the formation of toroidal particles was less pronounced than in the case of PSS as an organic additive. We further investigated the effect of DS concentration on the $CaCO_3$ particles formation (**Figure S5**). The obtained SEM images confirmed that the morphology of the obtained particles did not significantly changed depending on the DS concentration (0.5-2 mg/mL). Thus, the differently shaped $CaCO_3$ particles growth can be controlled by the employment of organic additives.

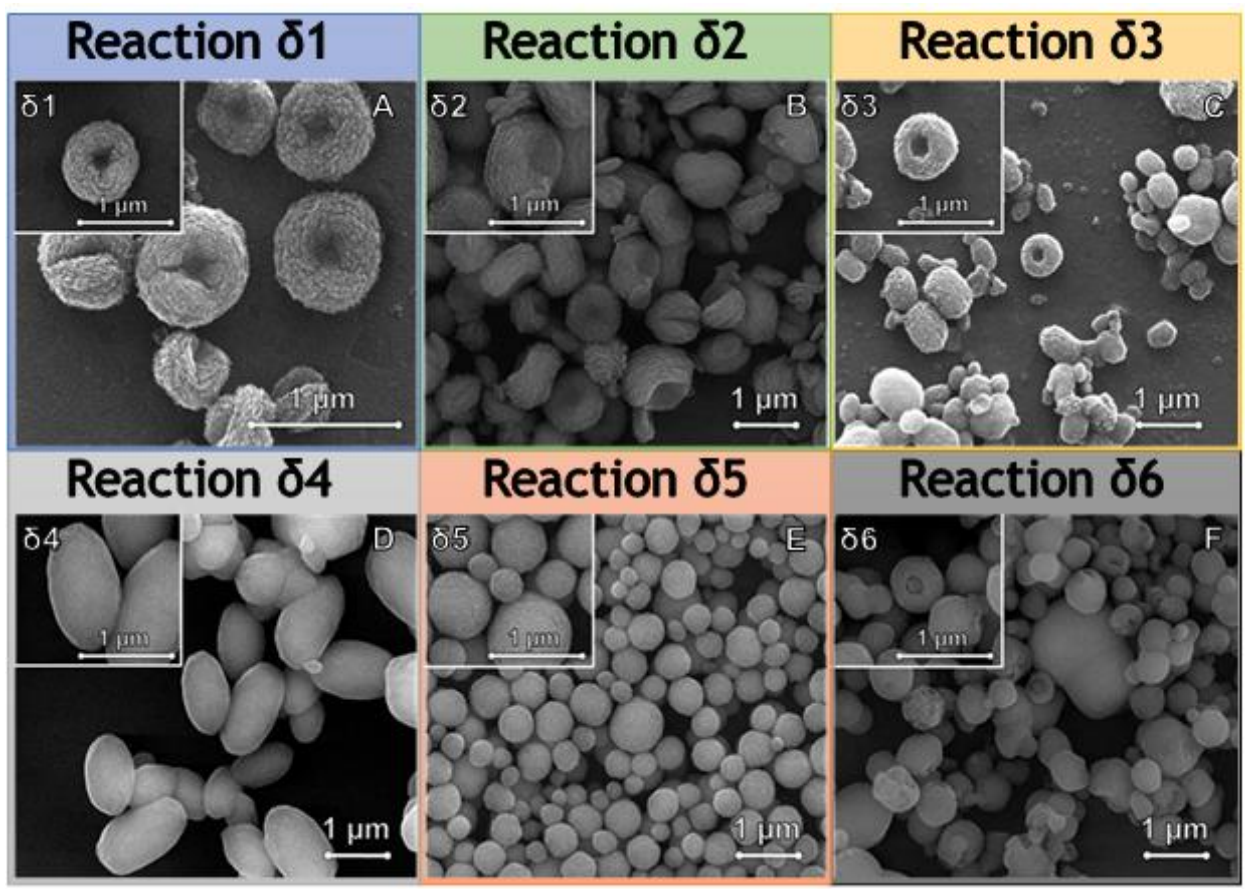

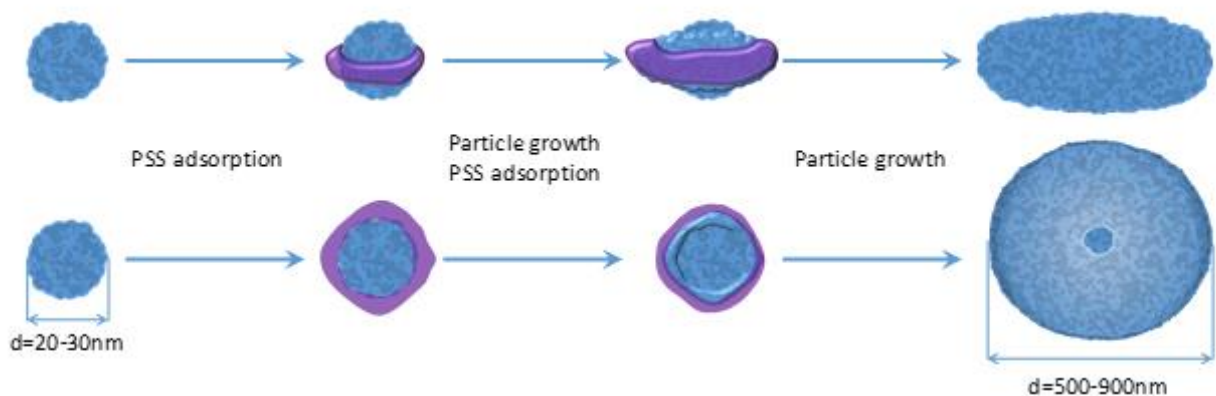

*Figure 6. Influence of organic additives on CaCO$_3$ particles formation: A. SEM images of the products obtained from the **reactions δ1- δ6** with corresponding schematic illustration of toroids formation by introducing PSS as organic additives.*

Differently shaped CaCO$_3$ particles were additionally evaluated with XRD analysis (**Figure 7**). The obtained data revealed that by the addition of 2 mg/mL PSS into the co-precipitation reaction with the excess of Ca$^{2+}$ ions (**reaction δ3**) toroidal particles were formed, which contained 100% vaterite phase. While in the **reactions δ1 and δ2** the mixture of calcite ($R\underline{3}c$)[50] and vaterite (*P6$_3$/mmc*)[51] phases was detected (72% and 58%, respectively). CaCO$_3$ particles comprised of almost 100% vaterite were obtained also in the **reactions δ4, δ5.** The overall data indicate that the additives in the co-precipitation reaction promote the formation of vaterite phase. These results are in agreement with the previously published works.[52,53] Indeed, it has been reported that the addition of organic polymers in the co-precipitation reaction, stabilize the amorphous $CaCO_3$ and prevent the and suppress the mineralization of $CaCO_3$.[54] We further calculated the sizes of crystallites of the obtained differently shaped $CaCO_3$ particles from the full-width at half-maxima of the XRD peaks with Rietveld refinement method (using FullProf program and Scherrer equation).[54–56] The formed micrometric sized particles consisted of nanosized crystallite aggregates. The crystallite sizes, as well as lattice volumes are enlistet in the Table S2. The size of crystalline particles ranges from 8.5 to 20 nm, it results in differences of $CaCO_3$ particles surface area and porosity obtained from the different reactions. With the increasing concentration of PSS, the crystallite size increase as well. The possible explanation for this can be the higher amount of the formed Ca-PSS globules resulting in the increased crystallite sizes. Interestingly, due the size effects, reduction of the crystal lattice volume for the crystallites with the smaller sizes can be observed.

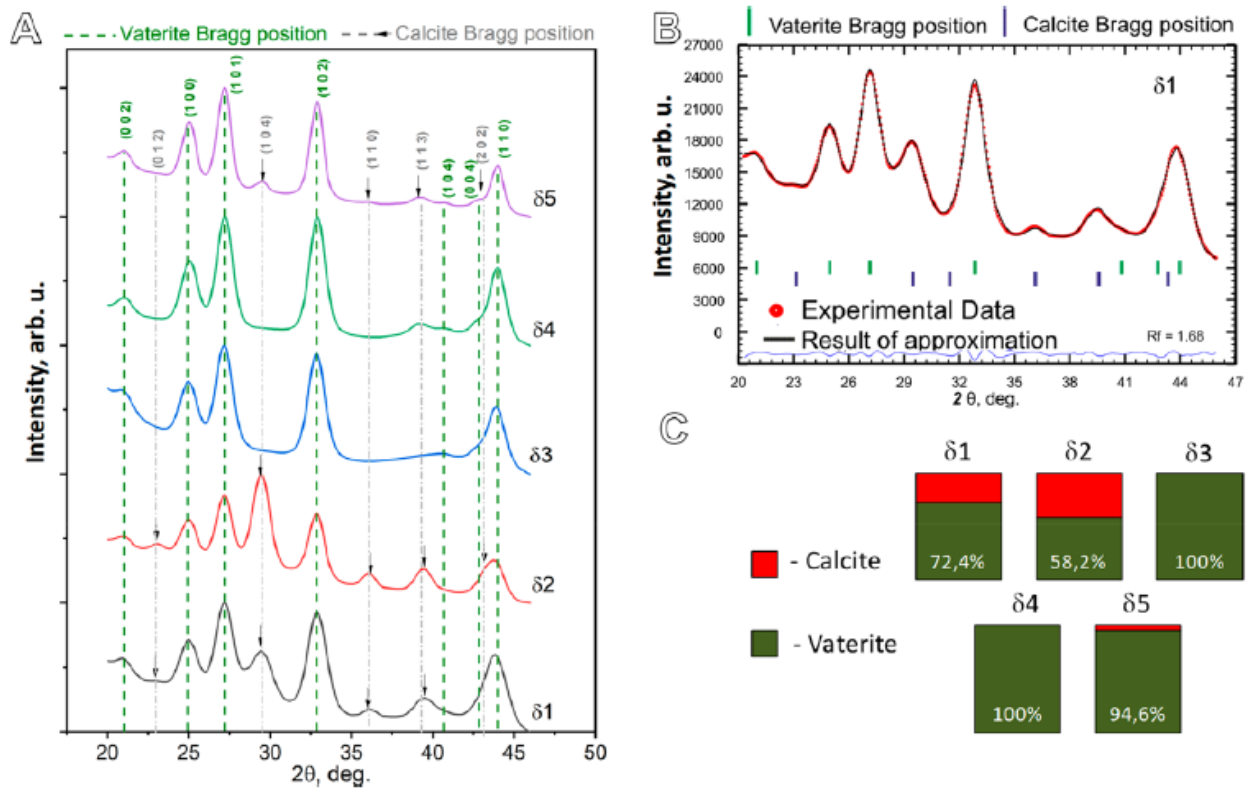

*Figure 7: XRD analysis A. X-ray powder diffraction patterns of $CaCO_3$ samples obtained in the **reactions δ1-δ5**. B. Example of Rietveld fit approximation of the powder XRD pattern from $CaCO_3$ particles obtained in the **reaction δ1**, where blue bars are calcite Bragg positions, green bars are vaterite Bragg positions, red circles are experimental data and black line is result of approximation. C. Phase content of $CaCO_3$ samples obtained in the **reactions δ1-δ5**.*

To summarize all the obtained results on CaCO$_3$ particles formation, we can distinguish three types of syntheses (**reactions β1, β3, δ1**), which can lead to the formation of stable CaCO$_3$ particles of different shapes with comparable sizes: spheroids, ellipsoids and toroids. Therefore, these differently shaped CaCO$_3$ particles were used for further experiments (**Figure S9**).

*3.3 Thermogravimetric analysis and thermal behavior of differently shaped CaCO$_3$ particles*

Next, we investigated thermal decomposition process of spheroidal, toroidal and ellipsoidal CaCO$_3$ particles (**reactions β1, β3, δ1**) by thermogravimetric analysis (TGA), and their thermal behavior was also analyzed by differential scanning calorimetry (DSC). The TGA curves of the samples are presented in **Figure 8A**. According to the results, it is clear that the decomposition process of spheroidal and ellipsoidal CaCO$_3$ particles (synthesis without organic additives) was similar, while the thermal destruction of toroidal CaCO$_3$ particles (synthesis in the presence of 0.5 mg/mL PSS) differed from spheroidal and ellipsoidal CaCO$_3$ particles. The one slight weight loss at 45-120 °C for all three samples was associated with the loss (removal) of physically adsorbed water from CaCO$_3$. Further, at 580-750 °C spheroidal and ellipsoidal CaCO$_3$ particles continued to lose the weight, what can be associated with the decomposition (calcination) of CaCO$_3$ to CaO and CO$_2$ (54% of weight loss for spheroids and 56 % for ellipsoids). Decomposition of both spheroids and ellipsoids started at around 580 °C. However, in case of spheroids the loss weight finished at 690 °C, while for ellipsoids it stopped at 720 °C. The decomposition process of toroids contained additional two phases at 410-520 °C and 550-650 °C, which are mostly due to the thermal degradation of PSS (organic additive) in CaCO$_3$ structure. Indeed, the TGA curve of pure PSS has similar trends in the weight loss between 400 and 650 °C (**Figure S10**). Due to the presence of organic additive, the final weight loss of toroidal particles containing PSS was 48%.

The influence of synthetic conditions and organic additive (PSS) on the thermal behavior of the differently shaped CaCO$_3$ particles was additionally studied with DSC analysis (**Figure 8B**). The DSC curves were measured in the range from 30 to 600 °C. According to the obtained data, spheroidal and ellipsoidal CaCO$_3$ particles showed two distinct thermal transitions. The first broad endothermic transition is assigned to the release of physically adsorbed water from the particles.[57] A first broad peak with maximum at around 100–120 °C was observed for all three types of particles (spheroidal, ellipsoidal and toroidal) and was more pronounced for the toroidal CaCO$_3$ particles. This can be attributed to the larger amount of water adsorbed onto the toroidal particles, which can be due to the presence of PSS. The exothermic process at 530 °C for spheroids and 420 °C for ellipsoids corresponds to transformation of disordered amorphous CaCO$_3$ into ordered CaCO$_3$ crystals (crystallization).[58] The difference in temperature of exothermic peaks for spheroidal and ellipsoidal CaCO₃ particles can be associated with differences in the size and shape of the particles, what is in agreement with previous work.[59] Indeed, Zou et al. demonstrated the importance of particle size on the thermal stability and crystallization of amorphous calcium carbonate.[59] Interestingly, the exothermic peak, which corresponds to crystallization process of CaCO₃, was missed in case of toroidal particles (synthesis was performed in the presence of organic additive – PSS) at the same temperature range as for spheroids and ellipsoids. This result indicates the inhibition effect of PSS on the crystallization of calcium carbonate under thermal conditions. Xu-Rong et al. showed the same tendency, where PSS inhibits the crystallization of calcium carbonate under thermal conditions and shifts the exothermic peak to the temperatures higher than 600 ˚C.[60]

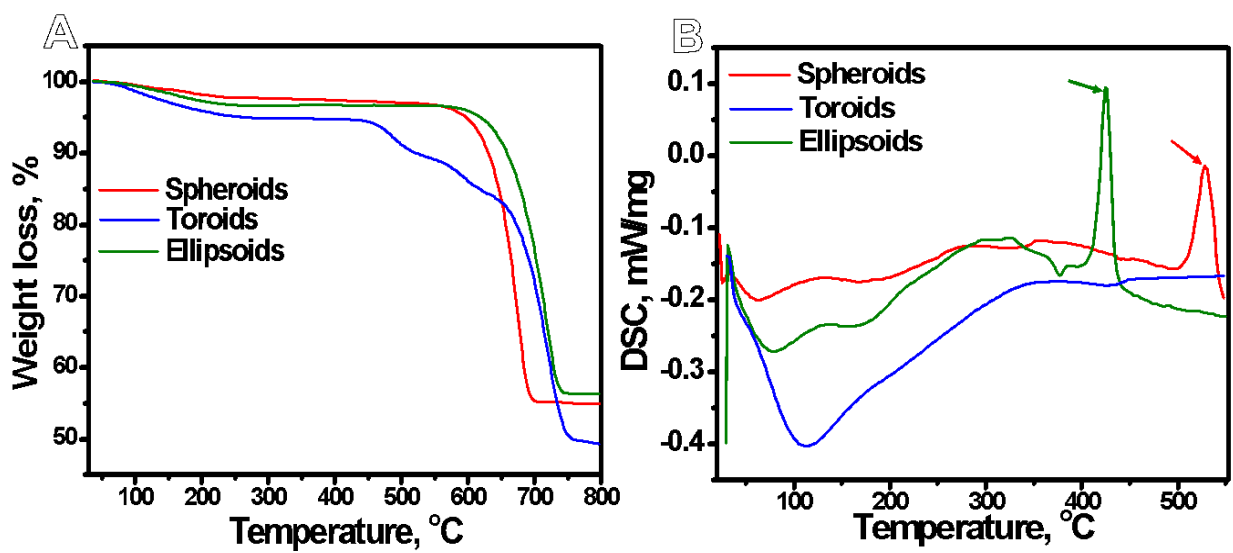

**Figure 8.** *Thermogravimetric analysis and thermal behavior of differently shaped CaCO₃ particles:* TGA curves (**A**) and DSC curves (**B**) of spherical (red), toroidal (blue) and ellipsoidal (green) CaCO₃ particles.

## 3.4 Surface area characterization, the adsorption capacity of differently shaped CaCO₃ particles and release efficiency

The obtained differently shaped CaCO₃ particles can be employed as carriers for delivery of bioactive compounds, therefore, the information about specific surface area is of highly importance. The nitrogen isotherms with the corresponding pore-size distribution of differently shaped CaCO₃ particles (**reactions β1**-spheroids**, β3**-ellipsoids**, δ1**-toroids) are presented in **Figure 9 A, B, C**. According to the IUPAC classification,[61] the spheroids and ellipsoids correspond to the type V adsorption/desorption profiles, while toroids to the type IV. The hysteresis loops were observed for all types of particles. The hysteresis loop is usually associated with the filling and emptying of the mesopores by capillary condensation. The form of hysteresis loops for ellipsoidal and spheroidal CaCO$_3$ particles corresponds to the type H1 (IUPAC classification).[61] In this case, particles exhibit a narrow range of uniform mesopores and network effects are minimal. In contrast to the hysteresis loops of type H2, which corresponds to the toroidal particles. Type H2 is characterized by more complex pore structures in which network effects are important.

The BET surface areas of all samples along with their total pore volumes and average pore sizes calculated with the BJH methods are listed in Table 1. The obtained results show that the spheroids are characterized by the lowest specific surface area (40.6 m$^2$/g) compared to ellipsoids (43.1 m$^2$/g) and toroids (75.5 m$^2$/g). The increased surface area of the toroidal particles can be associated with the inclusion of PSS into CaCO$_3$ particles, which can locate on the surface of particles and contributes to their porosity. The organic additive (PSS) can also be a reason of the decreased average pore size of toroidal particles (3.57 nm) compared to the spheroidal and ellipsoidal particles (12.7 and 15.7 nm, respectively), what can be explained by the filling of pores by the PSS resulting in the reduction of the average pore size.

Next, we investigated the adsorption properties ($Q_e$) of such particles using physical adsorption approach, which is widely used for the loading of sensitive biomolecules.[15] This is essential for further application of the developed carries as drug delivery carriers, especially, in the case of low molecular weight cargo, since the most commercially available drugs are small molecules (under 1000 Da). The effective method to encapsulate small cargo in proper drug carriers is important. To estimate the adsorption efficiency of low molecular weight dye (TRITC), the same mass of spheroidal, toroidal and ellipsoidal CaCO$_3$ particles (**reactions β1, β3, δ1**) was incubated with TRITC at different concentrations ranging from 0.5 mg/mL to 0.03 mg/mL. The adsorption isotherm curves were then plotted (**Figure 9G**). The corresponding CLSM images of obtained differently shaped particles with adsorbed TRITC are depicted in **Figures 9 D, E, F**. The Langmuir fitting was used to analyze the experimental adsorption isotherms, describing with following equation (Ref.[62]):

$$Q_e = \frac{Q_{max}bC}{1 + bC}$$

where b is the Langmuir isotherm constant, $Q_{max}$ is the theoretical monolayer saturation capacity of the TRITC, C is the equilibrium concentration of TRITC. The detailed description of calculation of adsorption capacity of differently shaped CaCO3 particles is presented in ***Supporting Information*** §7.

From the Langmuir isotherms it can be seen that already at the smallest concentration of TRITC, the loading capacity of toroidal $CaCO_3$ particles is higher than of ellipsoidal and spheroidal particles (**Figure 9H**). At the maximum of added adsorbent concentration (0.5 mg/mL), the adsorption capacity of toroids is 1.2 times higher than adsorption capacity of ellipsoids and 3 times higher than adsorption capacity of spheroids. This can be attributed with the surface properties of the resulted particles, what is in agreement with the Ref.[63] The adsorption of the TRITC onto the porous $CaCO_3$ particles occurs mostly due to the nonspecific interactions of dye molecules with the $CaCO_3$ porous surface.[64] Moreover, electrostatic forces are also involved in the adsorption process, since the TRITC molecules are positively charged. The negatively charged PSS embedded into the structure of toroidal particles (**reaction δ1**) eventually contributes to the more negative zeta-potential (-12 mV) compared to the spheroids and ellipsoids (-2 mV and -0.8 mV). Therefore, the positively charged TRITC adsorbed onto the surface of toroids in the higher rate due to the electrostatic interactions.[46] Moreover, toroids possess the higher surface area compared to the spheroids and ellipsoids, what can also affect the enhanced adsorption capacity of these particles.[63]

To evaluate the release abilities of differently shaped TRITC labeled $CaCO_3$ particles (spheroidal, toroidal and ellipsoidal), they were shaken in water for 24 h. At the definite time point (1, 2, 4, 6, 8, 20 and 24 h), the amount of released cargo was measured and the percentage of released TRITC was plotted versus time of incubation (**Figure 9I**). The release of TRITC from toroids demonstrated slightly lower release rate than from ellipsoids, whereas spheroidal particles showed the lowest. The release kinetics for all three types of particles was approximately the same: there is a clear deceleration of the cargo release after 10 h of shaking.

Desorption (release) of TRITC molecules from the $CaCO_3$ particles occurs due to the possible recrystallization of vaterite into calcite in the aqueous solution.[23] As it has been shown earlier in Ref.[23], exposure of non-modified vaterite to water media results in complete transition of vaterite into calcite, which possess a lower surface area. The obtained results confirm that the differently shaped $CaCO_3$ particles are able to adsorb and release bioactive molecules for at least 24 h.

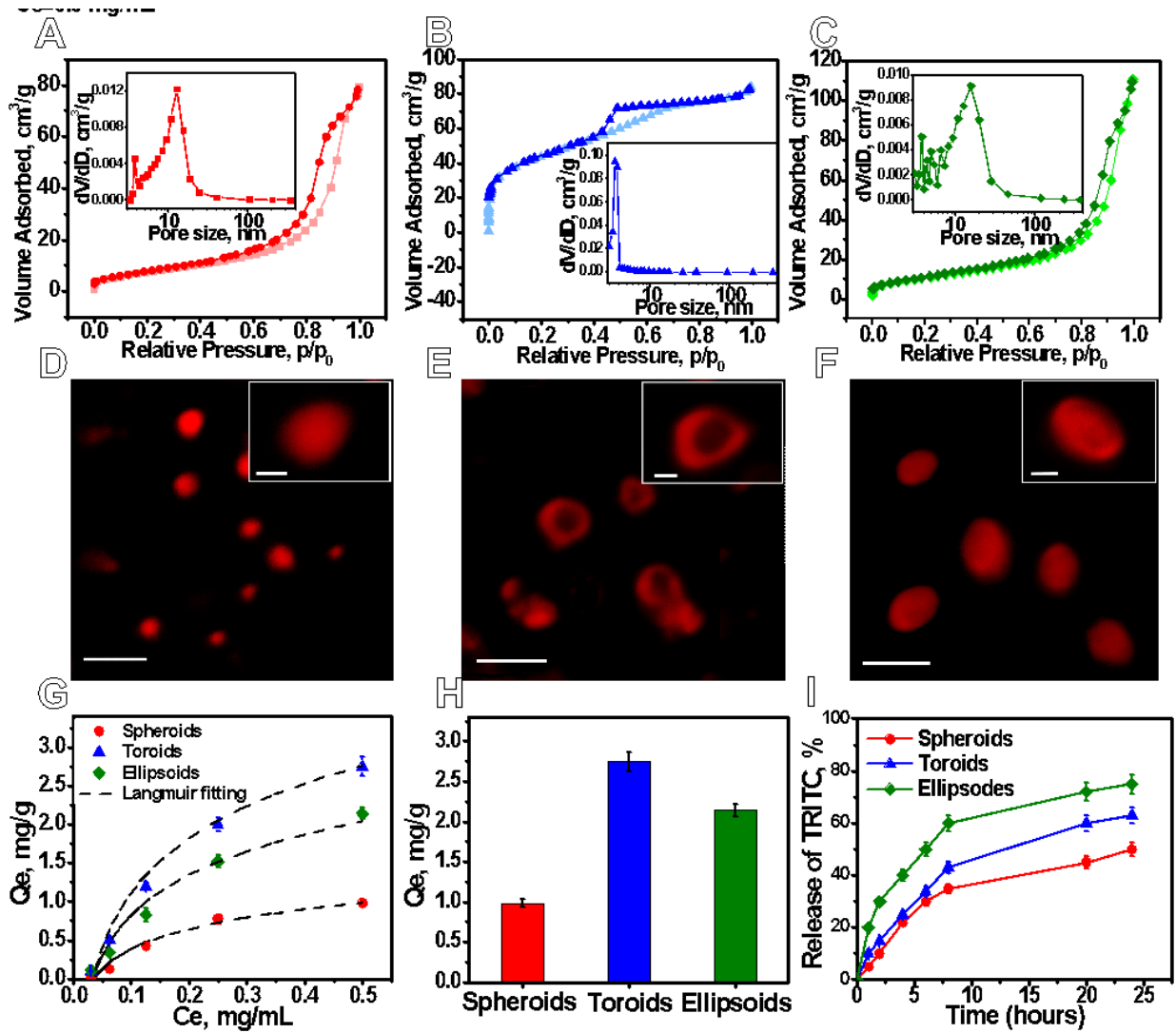

*Figure 9. Surface properties of differently shaped CaCO$_3$ particles with the corresponding adsorption/release efficiency:* Nitrogen adsorption - desorption isotherms and corresponding pore-size distribution curves (inserts): spheroids (**A**), toroids (**B**), ellipsoids (**C**). CLSM images of TRITC labeled CaCO$_3$ particles: spheroids (**D**), toroids (**E**), ellipsoids (**F**). Scale bars: **D**, **E**, **F** are 1 µm, inserts 200 nm (data are presented as mean ± standard deviation). **G.** Non-linear fits of adsorption isotherm curves for TRITC adsorption of spherical (red), toroidal (blue) and ellipsoidal (green) CaCO$_3$ particles. **H.** Adsorption capacity of spherical (red), toroidal (blue) and ellipsoidal (green) CaCO$_3$ particles incubated with TRITC at concentration Ce=0.5 mg/mL. **I.** Percentage of released TRITC from spherical (red), toroidal (blue) and ellipsoidal (green) CaCO$_3$ particles shaken in water over time (0-24 h).

*Table 1. Surface properties of differently shaped CaCO$_3$ particles*

|  | Surface area, m$^2$/g | Pore volume, cm$^3$/g | Pore diameter, nm | Zeta-Potential, mV |
|---|---|---|---|---|
| **Spheroids** | 40.6 | 0.127 | 12.7 | -2 |
| **Toroids** | 75.5 | 0.084 | 3.57 | -12 |
| **Ellipsoids** | 43.1 | 0.172 | 15.7 | -0.8 |

*3.5 Particle-dependent uptake by C6 glioma cells*

The influence of particle geometry on cellular uptake is highly important for the further consideration of such carriers in drug delivery. The number of internalized particles per cell was evaluated after 24 h of incubation by counting the particles inside cells via fluorescence microscopy and subsequent image analysis. C6 glioma cells were used as a model of tumor to study cellular uptake efficiency. Incubation period of 24 h was used since differently shaped particles (**reactions β1, β3, δ1**) may have different internalization kinetics, as well as different endocytic processes can proceed differently. Therefore, a longer incubation period ensured that the particle shape was only one parameter influencing the efficiency of uptake.[35] The cytoskeleton of cells was fluorescently stained using phalloidin conjugated with AlexaFluor488 (AF488) and cell nuclei with Propidium iodine (PI). The co-localization of TRITC labeled particles within the cellular compartments was examined with Z-stack option. **Figure 10A** shows the CLSM images of C6 glioma cells that had internalized the differently shaped particles. **Figure 10B** depicts three-dimensional CLSM image of a cell scanned from the glass slide to the top of the cell at a series of depth along the z-axis. The main panel of the **Figure 10B** shows the fluorescent image in the x-y cross section at a given z-location. The two smaller panels reveal the side views of cell along the x-z (top panel) and y-z cross sections (right side panel). It can be seen that the TRITC labeled particle (red) is located in-between two layers of fluorescently labeled cell membrane (green). 3D reconstruction of the cell was additionally performed using conventional ZEN 2009 software (**Figure 10B right**). As shown in **Figure 10B**, red-stained particles clustered near the cell nucleus in coplanar position. From the obtained Z-stack images, the frequency f(x) of cells which have internalized x particles per cell and corresponding cumulative probability p(x) were calculated and depicted in the histograms (**Figure 10C, D**). As shown in **Figure 10C, D**, and the toroidal particles were internalized to lower extent than spheroidal and ellipsoidal particles, while ellipsoidal particles were taken up at significantly higher amount with C6 glioma cells, what is in agreement with previous studies.[35] The increased penetration abilities of ellipsoidal particles can be associated with following reasons. First of all, the studied particles were mainly internalized with C6 cells via endocytosis[44,65] and depending on the particles shape the endocytosis efficiency was different. The contact angle between the particles and cell membrane, as well as the local curvature of the particle at the contact point with cell play a key role in the internalization efficiency of particles. In the case of ellipsoidal particles, the contact angle between particles and cell membrane is lower than for spheroidal and toroidal particles. Therefore, the internalization process of ellipsoids apparently requires less energy of vesicles that wraps the particles resulting in more efficient uptake.[66] Moreover, the decreased local mean curvature at the side edge of ellipsoidal particles favors the particles uptake.[67] The obtained results are of the great importance for the developing of the shape specific targeting carriers. For example, Kolhar et al. demonstrated that rod-shaped particles showed the greater accumulation in the organs compared to spheres in vivo.[68]

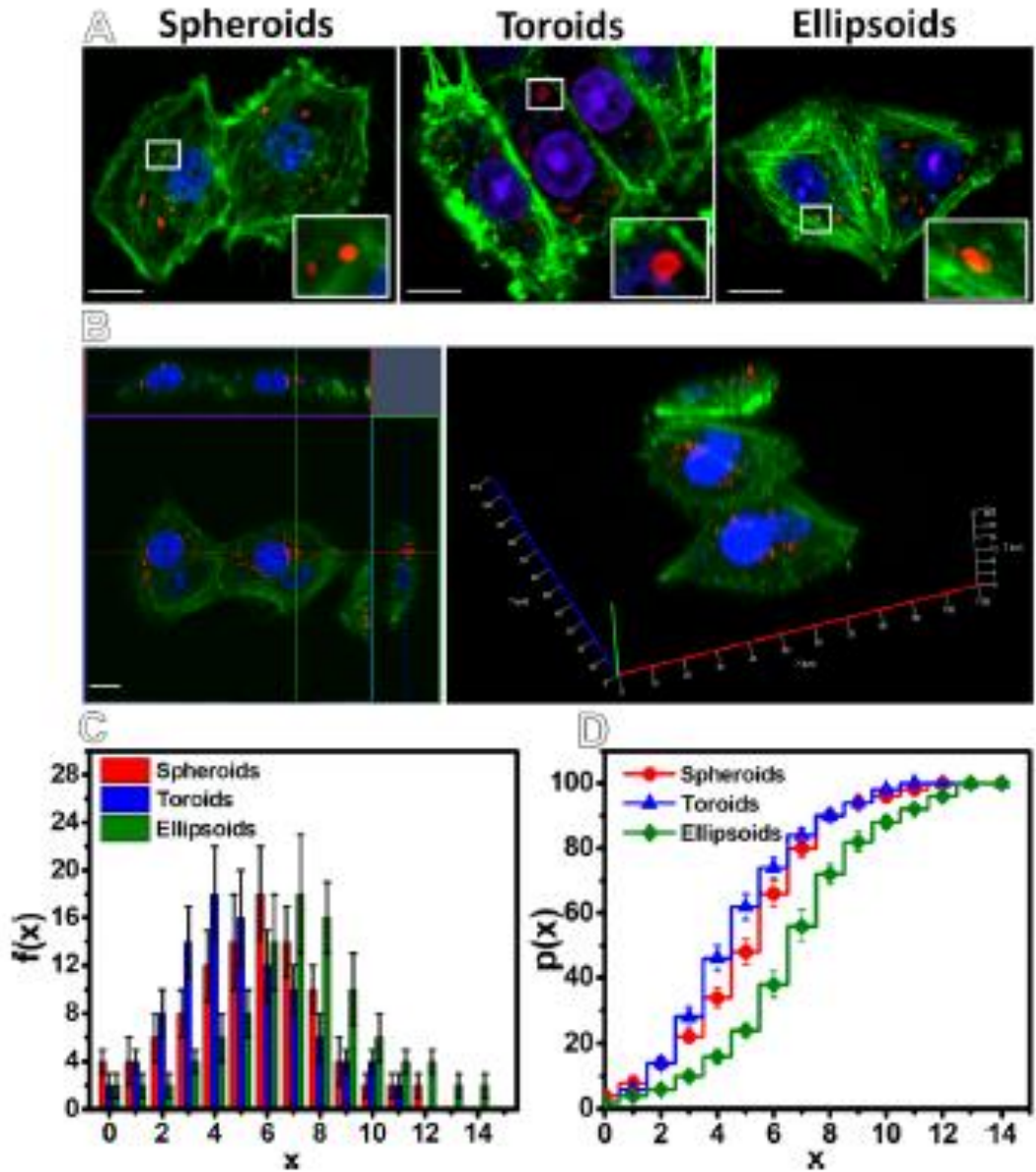

*Figure 10. Influence of particle shape on the internalization efficiency: **A.** CLSM images of C6 glioma cells with internalized particles labeled with TRITC added at cell-to-particle ratio = 1:10. Nuclei: blue stained with PI; Membrane: green stained with phalloidin-AF488; red stained $CaCO_3$ particles labeled with TRITC. **B.** Z-stack CLSM image confirming the internalization of spherical particles labelled with TRITC with corresponding 3D reconstruction image of cells with internalized particles. **C.** Histograms of the frequency f(x) of cells which have internalized x particles per cell after 24 h of incubation with particles at cell-to-particle ratio = 1:10. **D.** Corresponding cumulative probability plot p(x) for x internalized particles per cell. Scale bars: **A**, **B**, 10 μm (data are presented as mean ± standard deviation).*

### 3.6 Evaluation of particle toxicity

The toxicity of differently shaped particles at various cell-to-particle ratios of 1:6, 1:12, 1:25, 1:50 and 1:100 was evaluated on C6 glioma cell model as well (**Figure 11**). For this, differently shaped particles were incubated with cells for 24 h and then LIVE/DEAD assay was performed. As shown in **Figure 11,** the viability experiments displayed the absence of cytotoxicity of differently shaped particles (spheroids, ellipsoids and toroids) even at the highest cell-to-particle ratio (1:100). The obtained toxicity data are in agreement with the previously published data measured on HeLa carcinoma cells.[6] Zhang et al. evaluated the biological safety of porous $CaCO_3$ particles on HeLa cells and demonstrated biocompatibility of $CaCO_3$ particles in the micrometer and nanometer ranges.[6] Calcium is the key element in cell metabolism. Cells possess the mechanisms of tolerance to calcium, as well as effective calcium efflux through calcium channels. Therefore, while the fluctuations of calcium ions can influence the cell functions (adhesion, moving, signaling, action potential transduction), the cytotoxic action of sustained released $Ca^{2+}$ on cells is not observed.

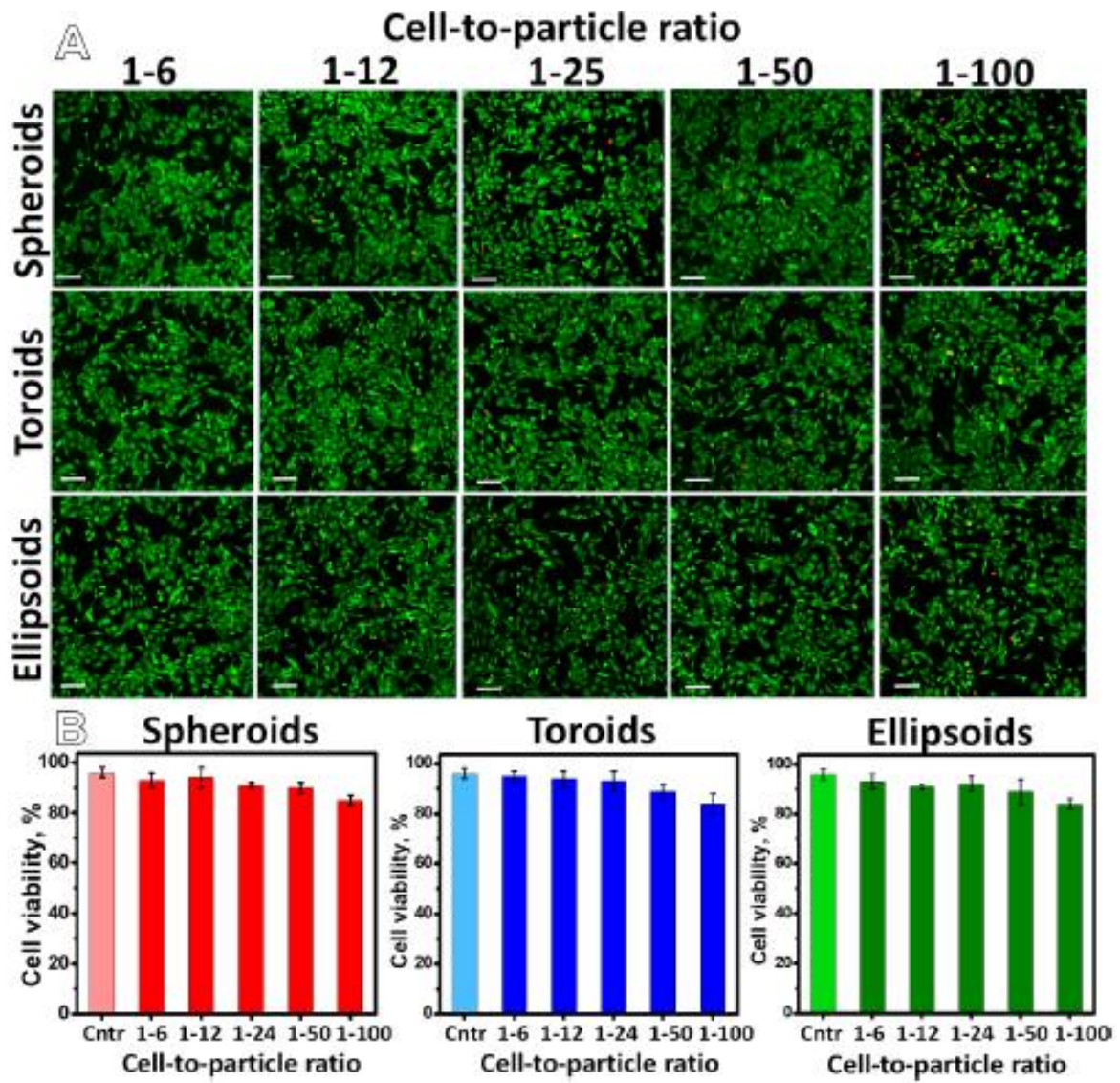

*Figure 11. Cytotoxicity of differently shaped $CaCO_3$ particles: A. CLSM images of C6 glioma cells incubated with differently shaped $CaCO_3$ particles at various cell-to-particle ratio for 24 h.*

*Living cells were stained with Calcein AM (green) and dead cells with PI (red).* ***B****. C6 glioma cells viability after incubation with differently shaped CaCO$_3$ particles. Scale bars: **A**, **B**, 100 μm (data are presented as mean ± standard deviation).*

**Conclusions**

We performed comprehensive study describing how different physicochemical parameters such as salt concentrations, salt ratios, times of the reactions, organic additives influenced the formation of CaCO$_3$ particles of different geometries and sizes. The obtained data revealed that the crystallinity of the resulted CaCO$_3$ particles depends on the reaction conditions and plays important role in the surface properties of CaCO$_3$ particles, colloidal behavior and phase-stability of particles. Such variations of the reaction conditions (**reactions β1, β3, δ1**) allows fabrication of differently shaped CaCO$_3$ particles (spheroids, ellipsoids and toroids) for further investigation of loading capacity and interaction with cells. Changes in the CaCO$_3$ particles anisotropy leads to the different adsorption capacities of low molecular weight bioactive compounds. In particular, toroidal shaped particles are characterized by the highest adsorption capacity compared to spheroids and ellipsoids. The solubility of the drug delivery carriers should be also taken into account. As shown in the recent study, CaCO$_3$ particles with the dominant vaterite phase possess increased solubility in the biological fluids.[69] Owing outstanding surface properties of CaCO$_3$ particles together with solubility enables their utilization as drug delivery carriers. Therefore, these particles were further used for studying the uptake efficiency on example of C6 glioma cells. As demonstrated, cell uptake depended on the shape of the CaCO$_3$ particles. Ellipsoids were internalized by C6 glioma cells in the highest rate. Moreover, the used differently shaped CaCO$_3$ particles did not show any cytotoxic effects even at 100 added particles per cells. As mentioned above, CaCO$_3$ particles are widely used as carriers for drug delivery due to its non-toxicity, biocompatibility and degradability in biological fluids. Synthesis of CaCO$_3$ particles allows obtaining carriers with different physicochemical properties (e.g. size, shape, morphology and so forth). It is worth noting that even uncoated vaterite particles, which are not stable in water, can be used as drug delivery carriers, since after the first contact of particles with the biological fluids, organic compounds tend to bind to the particles surface forming so-called "protein corona" around the particles. This corona can additionally stabilize the particles in biological fluids preventing its unwanted mineralization.[70] The developed differently shaped carriers can be considered as perspective drug delivery systems, which can eventually find their pathway as a convenient platform to carry various bioactive compounds.

**Supporting Information**

Other experimental results, describing synthesis of CaCO$_3$ particles and their characterization using SEM, XRD; procedure for adsorption capacity measurements and drug release study; cell studies (e.g. cell uptake, toxicity) are available in ***Supporting Information***.

**Acknowledgements**

The work related to the characterization and biological studies of delivery carriers was supported by the Russian Science Foundation, grant no. 19-75-10010 (A.S.T). M.V.Z. thanks the President's Scholarship SP-1576.2018.4. Part of this work related to the synthesis of delivery carriers was supported by the Russian Science Foundation, grant no. 19-75-00039 (M.V.Z). The cultivation of C6 glioma cells and the cell viability experiments were partly supported by the grant of the Russian Foundation for Basic Research, No. 19-015-00098 (A.R.M.). This work was supported in part by ERC StG "In Motion" (802279) and the PAZY Foundation.